\documentclass[12pt]{article}
\usepackage[dvips]{graphicx}
\begin{document}

\title{Optical vortex coronagraphs on ground-based telescopes}

\author{Charles Jenkins \\Research School of Astronomy and Astrophysics,\\ Australian National University,
Cotter Road, Weston, ACT 2611,
\\Australia\\{\tt charles.jenkins@anu.edu.au}}

\maketitle

\begin{abstract}The optical vortex coronagraph is potentially a remarkably
effective device, at least for an ideal unobstructed telescope. Most
ground-based telescopes however suffer from central obscuration and
also have to operate through the aberrations of the turbulent
atmosphere.  This note analyzes the performance of the optical
vortex in these circumstances and compares to some other designs,
showing that it performs similarly in this situation. There is a
large  class of coronagraphs of this general type, and choosing
between them in particular applications depends on details of
performance at small off-axis distances and uniformity of response
in the focal plane. Issues of manufacturability to the necessary
tolerances are also likely to be important.

\end{abstract}

\section{Introduction}

\noindent Many areas of astrophysics need a coronagraph of very high
dynamic range, to detect faint sources near bright ones. Examples
include exoplanets, protoplanetary disks, and the structure of
active galactic nuclei.  These applications cover a wide range of
acceptable dynamic range, with planets being the most challenging.

A wide  variety of coronagraphs has been proposed\cite{guyonreview}
and one particularly impressive one is the optical vortex
coronagraph (OVC). This a remarkable example of diffraction at work
\cite{Foo,Lee,Palacios}. A ``vortex phase mask'' (VPM) is introduced
at the focus at the focus of a circularly symmetric optical system.
In the best-studied examples, this mask introduces a phase
proportional to $m \theta$, where $\theta$ is an azimuthal
co-ordinate in the focal plane, and $m$ is an even integer called
the topological charge. If the diffraction pattern at the focus is
an Airy function, then one finds exactly, i.e. analytically, that
all of the light in this diffraction pattern is diffracted outside
the re-imaged pupil. A simple Lyot stop will thus extinguish all of
the light from an on-axis source, at least for perfect optics.
Actual implementations of a VPM will have imperfections, including
possibly a loss of amplitude close to the vortex core, depending on
how the VPM is implemented. Here I concentrate on an ideal system.

The vortex phase mask appears to form the basis of a very elegant
coronagraph.  It has high performance by comparison with other
designs \cite{Guyon2}, and can be designed to be achromatic
\cite{Swartzlander} and insensitive to aberrations.
{\cite{Palacios2}  Its conception can be seen as a continuation  of
related designs -- the phase-mask coronagraph \cite{Roddier} and the
four-quadrant phase-mask coronagraph. \cite {Rouan}

It is of interest to examine the performance of the OVC on
ground-based telescopes.  These generally have two features which
are not favourable for the OVC - a central obstruction, and images
blurred by atmospheric turbulence.  I will examine both aspects and
show that the OVC behaves similarly to some other designs, but with
good performance and some  advantages even in these non-ideal
circumstances.

\section*{The non-ideal PSF -- the central obstruction}

For any coronagraph, the loss of Strehl ratio from a central
obstruction is a disadvantage.  The OVC is no exception;  in the
reimaged pupil, the secondary is surrounded by a halo of light,
whose total power is comparable to the power obstructed by the
secondary. This result is illustrated by direct computation in
Figure \ref{figure1}.

\begin{figure}[htb]
\center{\includegraphics[width=10cm]{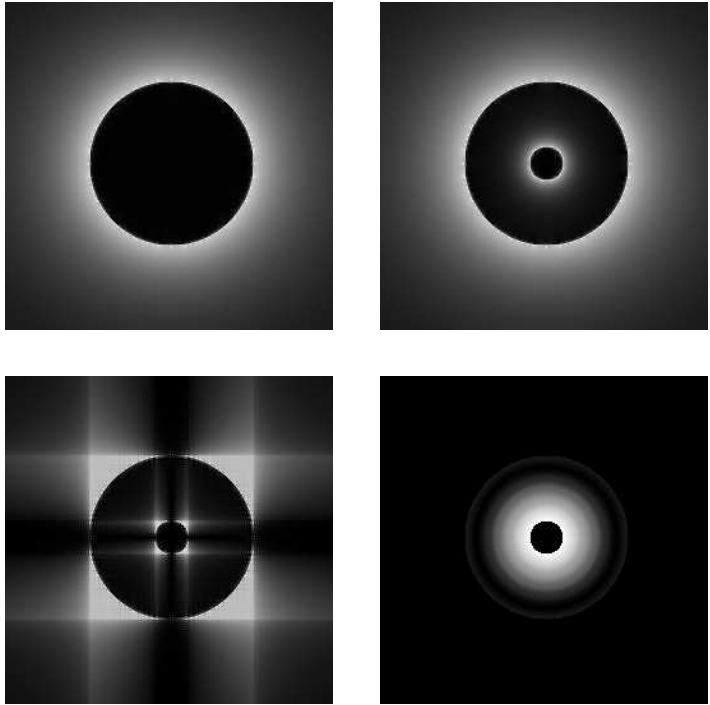}} \caption{The
intensity pattern in the reimaged pupil plane beyond a VPM, for an
obstructed and unobstructed pupil. This was computed by direct FFT
for perfect optics and charge 2 (top left), for charge 2 with a
central obstruction (top right), for a quadrant phase mask (bottom
left) and an optimum prolate apodization (bottom right).}
\label{figure1}
\end{figure}

This result can be examined analytically, as is discussed elsewhere
in detail by Mawet {\em et al.}\cite{mawet}. Only  the central
obstruction need be added to their comprehensive treatment. I
discuss a mask of charge 2 as an example. Using standard Fraunhofer
diffraction, the amplitude, in a reimaged pupil plane beyond the
VPM, is given by a generalized Hankel transform:

\[
{\cal A}(r')=K e^{2 \imath \beta} \int_0^{\infty} {\cal F}(k) \,
J_2(k r') k \, dk
\]
in which $r$ and $k$ are suitably scaled  radial co-ordinates in the
reimaged pupil and the preceding  focal plane, and $\beta$ is an
azimuthal angle in the reimaged pupil. Constants irrelevant to the
discussion are absorbed into $K$. ${\cal F}$ is the amplitude in the
focal plane for an on-axis source. If this is an Airy function for
an input pupil of radius $R$, the integral becomes

\[
{\cal A}(r)=K e^{2 \imath \beta} R \int_0^{\infty} J_1(k R) \, J_2(k
r') \, dk.
\]
(The second-order Bessel function arises because the amplitude
introduced by the VPM has been taken to be $e^{2 \imath \theta}$ in
this example.  This topological charge of 2, via the angular
integral of the Fourier transform, translates directly into the
order of the Bessel function.  The needed result is

\[ \int_0^{2\pi} e^{2 \imath \theta} e^{\imath k (\theta-\beta)} \, d
\theta = -2\pi e^{2 \imath \beta} J_2(k) \,)\] (see equation
\ref{generating}).

The infinite integral is a  case of  the Weber-Schafheitlin integral
\cite{watsona}, and so

\begin{equation}
\begin{array}{lcll}
{\cal A}(r') & = & 0 &   0 \leq r' < R \\
& & &  \\
 & = &  K e^{2 \imath \beta}(\frac{R}{r'} )^2 &  r' \geq R
\end{array}
\label{clear}
\end{equation}

for the amplitude in the reimaged pupil.  This is the result that
motivates the optical vortex coronagraph; similar results can be
obtained for other even topological charges of the VPM. The power
diffracted outside $R$ is $\vert K \vert^2 \pi R^2$, which is
exactly equal to the power in the focal plane diffraction pattern
$(J_1(k R)/k R)^2$. In the case of a centrally obstructed pupil,
however, an on-axis source at infinity produces an amplitude in the
focal plane that is given by

\[
{\cal F}(k) = K (\frac{R J_1(k R)}{k}-\frac{a J_1(k a)}{k})
\]
in which $a$ is the radius of the central obstruction.  It follows
that the amplitude in the reimaged pupil must be

\begin{equation}
\begin{array}{lcll}
{\cal A}(r') & = & 0 &    0 \leq r' < a \\
& \\
& = & -K e^{2 \imath \beta}(\frac{a}{r'} )^2 & a \geq r' < R \\
& \\
 & = &  K e^{2 \imath \beta}\left((\frac{R}{r'} )^2 -(\frac{a}{r'} )^2 \right) &  R \geq
 r'
\end{array}
\label{fullscatter}
\end{equation}

The total diffracted power is now $\vert K \vert^2 \pi (R^2-a^2)$,
and the power diffracted into the reimaged pupil between $a$ and $R$
is $\vert K \vert^2 \pi a^2 (1-(a/R)^2)$.   The total energy
diffracted into the halo of the secondary is nearly equal to the
energy incident upon it, particularly for typical telescopes where
$a$ is much less than $R$.

A first-order approach to evaluating this issue is simply to examine
the total power which passes through a  Lyot stop. In fact, one
might also consider blocking some of the diffracted light with an
enlarged stop at the secondary; while this costs light, it might
yield a net benefit in terms of attainable dynamic range.

\begin{figure}[htb]
\center{\includegraphics[width=10cm]{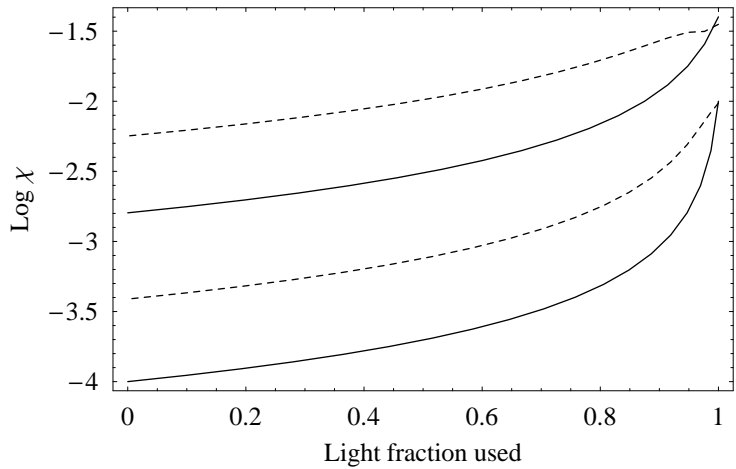}}
 \caption{The normalized diffracted power $\chi$,
 plotted against the fraction of the light that is admitted through a
 centrally-obstructed Lyot stop. The curves are for linear central obstructions $a/R$ of 0.2 and 0.1. The full line is
 for a VPM of charge 2, the dashed line for charge 4.}
 \label{figure2}
\end{figure}

Using Equation \ref{fullscatter}, the fraction of the power from an
on-axis source that appears between $\rho>a$ and $R$ is
\begin{eqnarray}
\chi & = & \left( \int \left (\frac{a}{r} \right)^4 2 \pi r \, dr
\right)/\left(\pi (R^2-\rho^2) \right)\nonumber \\
& = & \frac{a^4}{R^2 \rho^2}.
\end{eqnarray}
This is the fraction of the energy of the  the on-axis source that
passes the proposed Lyot stop. This may usefully be compared to the
fraction of the available light that is used  in this scheme, which
is
\[ \frac{R^2-\rho^2}{R^2-a^2} \]
as is shown in Figure \ref{figure2}. For modest loss of light
($\sim$ 20\%) the attainable dynamic range is about 100; this may be
useful, especially if the dynamic range is more fundamentally
limited by adaptive optics considerations.  Losing light is not
always a disaster if another parameter is improved, in this case
dynamic range, but it is clear that the sacrifice would have to be
very substantial to attain high dynamic range.

\begin{figure}[htb]
\center{\includegraphics[width=10cm]{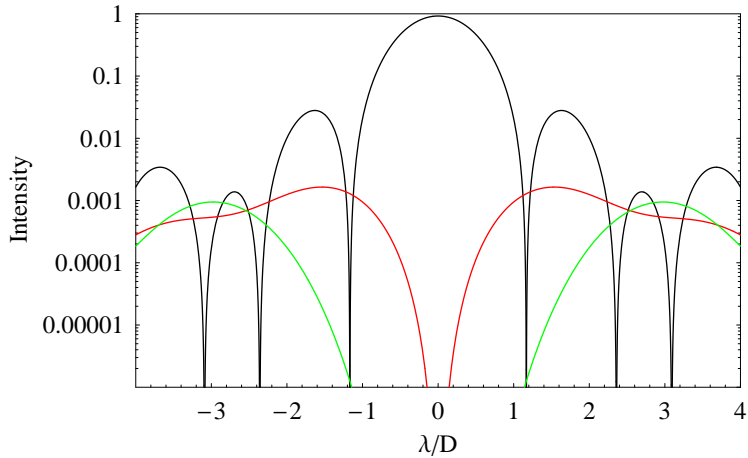}}
\caption{Calculations of the final point-spread function, for an
on-axis object.  The intensities are normalized to the
diffraction-limited case with no central obstruction. Colours code
the no-vortex case $m=0$ (black), charge  2 (red), and $m=6$
(green), all with a 20\% linear central obstruction.}\label{figure3}
\end{figure}

A fuller evaluation looks at the point-spread function of an on-axis
source, imaged through the VPM and a standard Lyot stop.  This
involves a Hankel transform of the amplitude in $ a \geq r' < R$, as
given by Equation \ref{fullscatter} (or its extension for higher
charges than 2).  This integral is again of the Weber-Schafheitlin
form.  For charge 2, for example, the final focal-plane amplitude,
in terms of a radial co-ordinate $k'$, is proportional to

\[ a^2 \left(  \frac{J_1(k'a)}{k'a} -\frac{J_1(k' R)}{k' R} \right). \]

Examples of the PSFs are shown in Figure \ref{figure3}, where it is
seen that the effects of the central obstruction are felt in
different parts of the image, depending on the charge on the VPM.
Depending on application, this level of light pollution may be
acceptable.

For a ground-based telescope, partial correction for atmospheric
turbulence may be  a more relevant limitation  to consider.
Anticipating the discussion of Section \ref{aocorrection}, Figure
\ref{figure4} shows that the effect of a central obstruction could
be quite perceptible. In this example the available ``dark'' region
is completely filled by diffracted light.

\begin{figure}[h]
\center{\includegraphics[width=0.6 \textwidth]{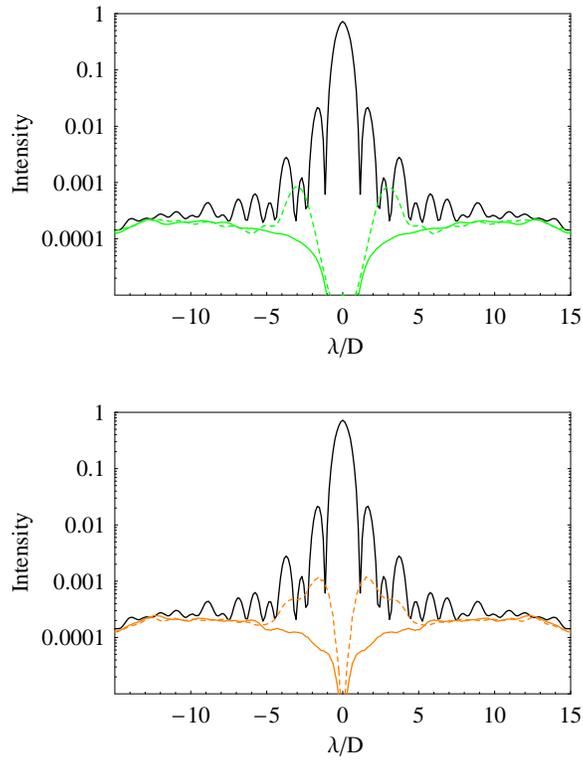}}
\caption{Simulation results, described later, for the on-axis
long-exposure point spread function in a well-corrected AO case. The
intensities are normalized to the diffraction-limited case. Colours
code the no-vortex case $m=0$ (black), charge  6 VPM (solid green),
and a quadrant phase mask (solid orange).  Dashed lines are for a
case with a 20\% linear central obstruction.}\label{figure4}
\end{figure}
\clearpage

 Several methods are possible to recover the performance
of the vortex coronagraph. For example, one might use a focal plane
occulting stop as a second stage. Because the images formed on-axis
through a VPM are quite wide (Figure \ref{figure4}) this stop would
have to be large.

Another option is to eliminate the central obstruction of the
telescope, perhaps by using an off-axis Herschel-like
system\cite{dooley}, as is proposed for the Terrestrial Planet
Finder mission. Large off-axis paraboloids are now being
manufactured for the Giant Magellan Telescope \cite{Johns}, which
will contain six adjacent mirrors in an outer ring. From the Fourier
shift theorem, the amplitudes in the reimaged pupils will simply be
displaced versions of Equation \ref{clear}, so that the issue will
become the light pollution between mirrors (Figure \ref{figure5}).
By using some subset of the available mirrors in a dilute aperture
configuration, light pollution may be substantially reduced,
although at the cost of losing light-gathering power.  This is an
example of how the central-obstruction problem may be handled with a
more dilute aperture.  Detailed calculation would be necessary to
establish that the light pollution was acceptable for a particular
configuration of apertures.  An additional option would be to
include stops close to the focal plane that prevented light
pollution from one aperture to the other, although this would limit
field of view.

\begin{figure}[h]
\center{\includegraphics[width=0.6 \textwidth]{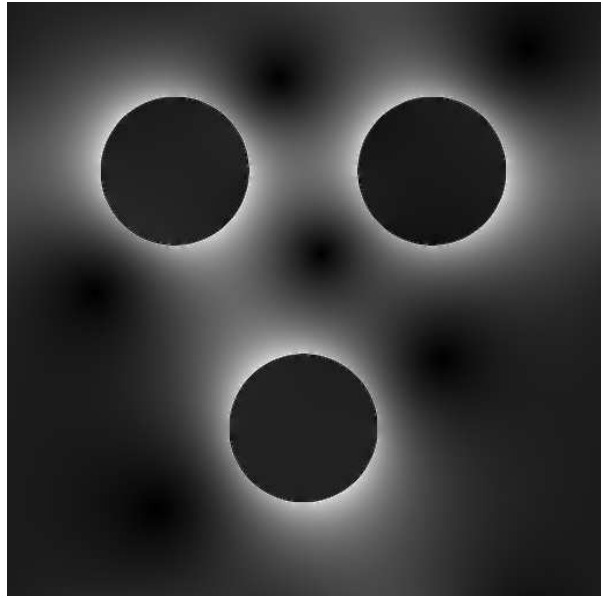}}
 \caption{The intensity pattern in the reimaged pupil
 plane beyond a VPM, for an illustrative telescope containing three off-axis mirrors.}
 \label{figure5}
\end{figure}

The most attractive approach seems to be Guyon's \cite{Guyon}
technique of phase-induced amplitude apodization. This reshapes the
illumination of the pupil, using the complex aspheric surfaces which
it is now possible to manufacture.  Guyon gives examples in which
the central obstruction os a two-mirror telescope is eliminated
completely. Guyon was also concerned to tailor the resulting
point-spread function so that it had low sidelobes, but there seems
no reason why his technique should not be used in a simpler way to
produce a pure Airy disk for an on-axis source. Apart from
manufacturability, which is still costly, the main limitation to the
method is the  narrow field of view.  If the pupil were reshaped to
eliminate the central obstruction, then an off-axis source will have
a phase discontinuity at the centre of the reshaped pupil. This will
limit the field to an angular extent given by the diffraction limit
of a pupil of equal size to the obstruction.  This will probably not
be a limitation however in the very narrow-field applications of
high-contrast coronography. It is also possible to avoid this
field-of-view limitation by a second stage of pupil reshaping after
the VPM, in which the original pupil is restored.\cite{Guyon}

\section{The non-ideal PSF -- partial AO correction}
\label{aocorrection}

A coronagraph on a ground-based telescope will probably be designed
to deal with a partially-corrected image, where the effects of
atmospheric turbulence have been partially removed. Such designs
have been made, for example, for the VLT \cite{VLTquad,VLTcorona}
and Gemini \cite{Geminicorona} telescopes.

In what follows I will assume that the central obstruction either
does not exist or has been removed by Guyon's method.  As Figure
\ref{figure4} shows, this is important.

The Point Spread Function (PSF) of an adaptive optics system is of
course affected by a large number of parameters, but when residual
phase errors are small the PSF is well approximated by a
diffraction-limited core, flanked by a halo whose functional form is
given by the power spectrum of the residual phase.  An optimized AO
system can largely remove low spatial frequencies from the
atmospheric phase power spectrum, but scales within or comparable to
actuator spacings cannot be removed.  The overall effect is that the
diffraction-limited part of the PSF sits in a ``hole'' in the
power-law halo.

Some analytical progress is possible in the case of well-corrected
images, and this provides a valuable guide to what may be expected
and a check in various cases of numerical simulations.  This is
useful in coronagraphic applications where a large dynamic range is
enquired in the calculation and small numerical errors may become
relatively important.  The essential results are the effect of a PVM
on a partially-corrected image, and some knowledge of the throughput
of these systems for targets that are not quite on-axis (the targets
of interest of course).

It is possible to calculate the form of the PSF in the limiting case
of good correction with high actuator density, for a coronagraph
built around a VPM.  The calculation is outlined in Appendix 1, and
shows that the coherent diffraction-limited core is simply removed
from the PSF, with only the halo (the residual phase power spectrum)
remaining.  This is intuitively what one would expect, since the VPM
is fundamentally an interference device.

In Appendix 2 it is shown that the throughput of a OVC, for an
off-axis source, is a strong power-law function of offset distance,
with the index depending on the charge of the mask.

Finally, in Appendix 3 it is shown that a wide variety of useful
phase masks are, in effect, superpositions of VPMs.

Taken together, these results suggest that the attainable dynamic
range of OVCs and related systems, in the partially-corrected AO
case, will converge at large radii to values fixed by the AO system
itself, while more marked variations in performance will be apparent
at small radii. To investigate this in more detail, three
general-purpose coronagraphic designs will be studied:  OVCs for
charges 2 and 6, a quadrant phase mask \cite{Rouan}, and a
prolate-apodized mask \cite{prolate}. The prolate mask affects
amplitude, not phase, and so gives excellent performance at the
expense of a loss of light. The  example used here is the
$\Lambda=0.99$ case reported by Soummer et al., which transmits only
26\% of the incident energy.

Calculating the performance of these systems in the case of partial
AO correction requires Monte Carlo simulation.   The technique used
here for numerical simulations is discussed in detail by
Sivaramakrishnan {\em et al.}.\cite{Sivaramakrishnan}. These authors
describe a useful way of summarizing  the AO system performance with
just  two parameters, the initial uncorrected turbulence strength
(the usual $D/r_0$) and an effective actuator spacing
$d_{\mbox{eff}}$. Two illustrative AO systems  are  modelled here to
investigate the performance of the OVC. The first is a fairly
high-performance AO system, with $D/r_0=10$ and
$D/d_{\mbox{eff}}=25$.  Here the diffraction-limited part of the PSF
is quite large (extending to about $8\lambda/D$.  The second models
a more general-purpose AO system, with $D/r_0=10$ and
$D/d_{\mbox{eff}}=10$.

Some care has to be taken in these numerical simulations as very
large arrays would be needed both to eliminate low-level aliasing
effects and also to give adequate detail in the computed PSFs.  For
the results to be discussed, an aperture of 128 pixels diameter was
embedded in a 1024 square array.  The results were checked
explicitly against analytical results, both the absolute accuracy
and for trends.

Aliasing is an issue for the high levels of accuracy that are
required in this type of simulation.  The VPM diffracts light
symmetrically  outside the pupil according to a power law, and this
slow decline aliases back into the pupil where there should be zero
amplitude.  The effect is small but perceptible.  It can be largely
eliminated by deriving the amplitude within the pupil for a
diffraction-limited source, and then subtracting this from
calculations in other cases.  One check on this method is to compare
analytical solutions for an obstructed telescope with the
simulations corrected in this way (see Figure \ref{figure6}) - the
method works well for the VPMs, giving reliable results at the
$10^6$ level of dynamic range.  The quadrant mask simulations work
well to at least this level without any corrections, probably
because of the strong angular modulation of extra-pupil diffracted
light (Figure \ref{figure1}).

\begin{figure}[htb]
\center{\includegraphics[width=10cm]{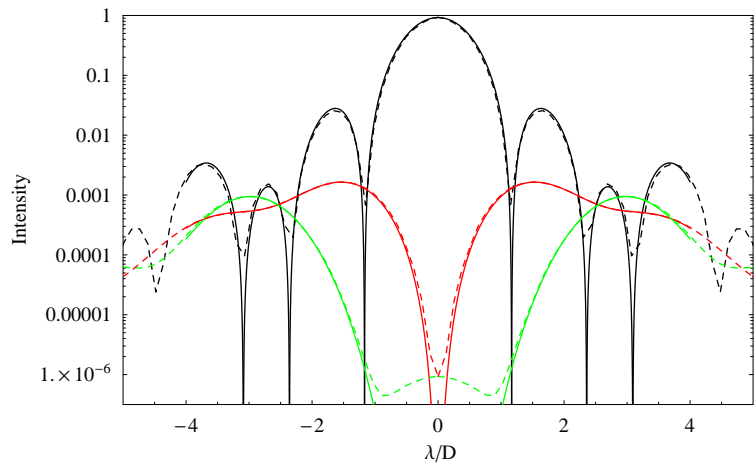}}
\caption{A comparison of analytical results for the PSF with
numerical calculations for a diffraction-limited  telescope with a
20\% linear central obstruction. Colours code the no-vortex case
$m=0$ (black), charges 2 (red) and 6 (green). Dashed lines are for
the numerical calculations, which are for a 128-pixel diameter
aperture embedded in a 1024-pixel square array, with an aliasing
correction as described in the text.}\label{figure6}
\end{figure}

The results are shown in Figures \ref{figure7} and \ref{figure8}.
The diffraction-limited ``core'' region is apparent, transitioning
into the power-law ``halo''.  The halo is unaffected by the
coronagraphs and the interesting region is within the ``shoulder''
that marks the transition. The extent of this region is determined
by the turbulence strength and the actuator density.

\begin{figure}[htb]
\centering{\includegraphics[width= .85\textwidth]{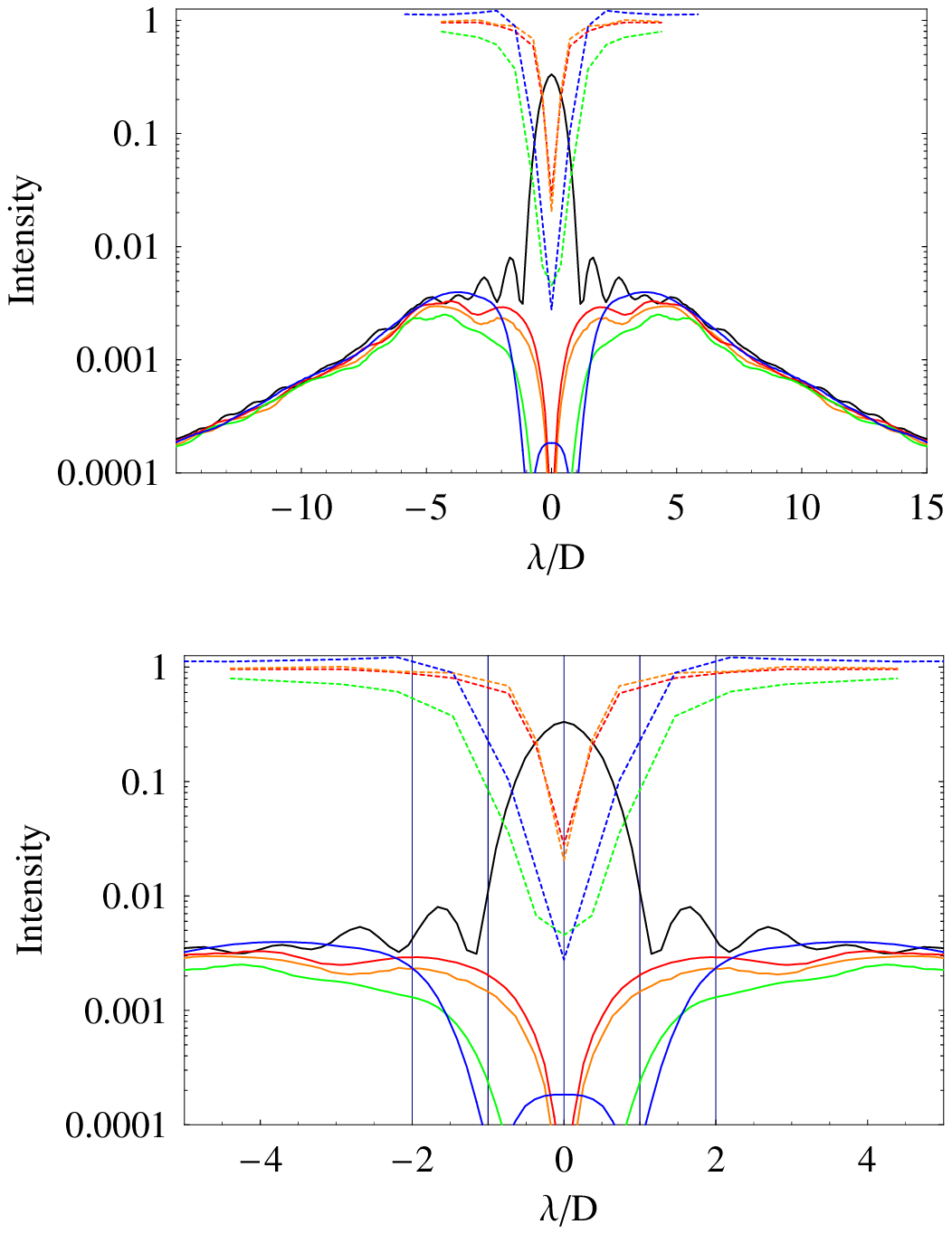}}
\caption{Simulation results for the long-exposure point spread
function of an on-axis source, for a the modestly-corrected AO case
(no central obstruction, $d_{\mbox{eff}}=r_0$). The intensities are
normalized to the diffraction-limited case. Colours code the
no-vortex case $m=0$ (black), charges 2 (red) and 6 (green), as well
as the quadrant phase mask (orange) and prolate-apodized mask
(blue). Dashed lines show the relative throughput for an off-axis
source, as a function of off-axis distance.  Calculations for the
quadrant phase mask, here and elsewhere, were at the optimum
$45^\circ$ to the lines of phase discontinuity.}\label{figure7}
\end{figure}
\clearpage
\begin{figure}[htb]
\centering{\includegraphics[width= .85\textwidth]{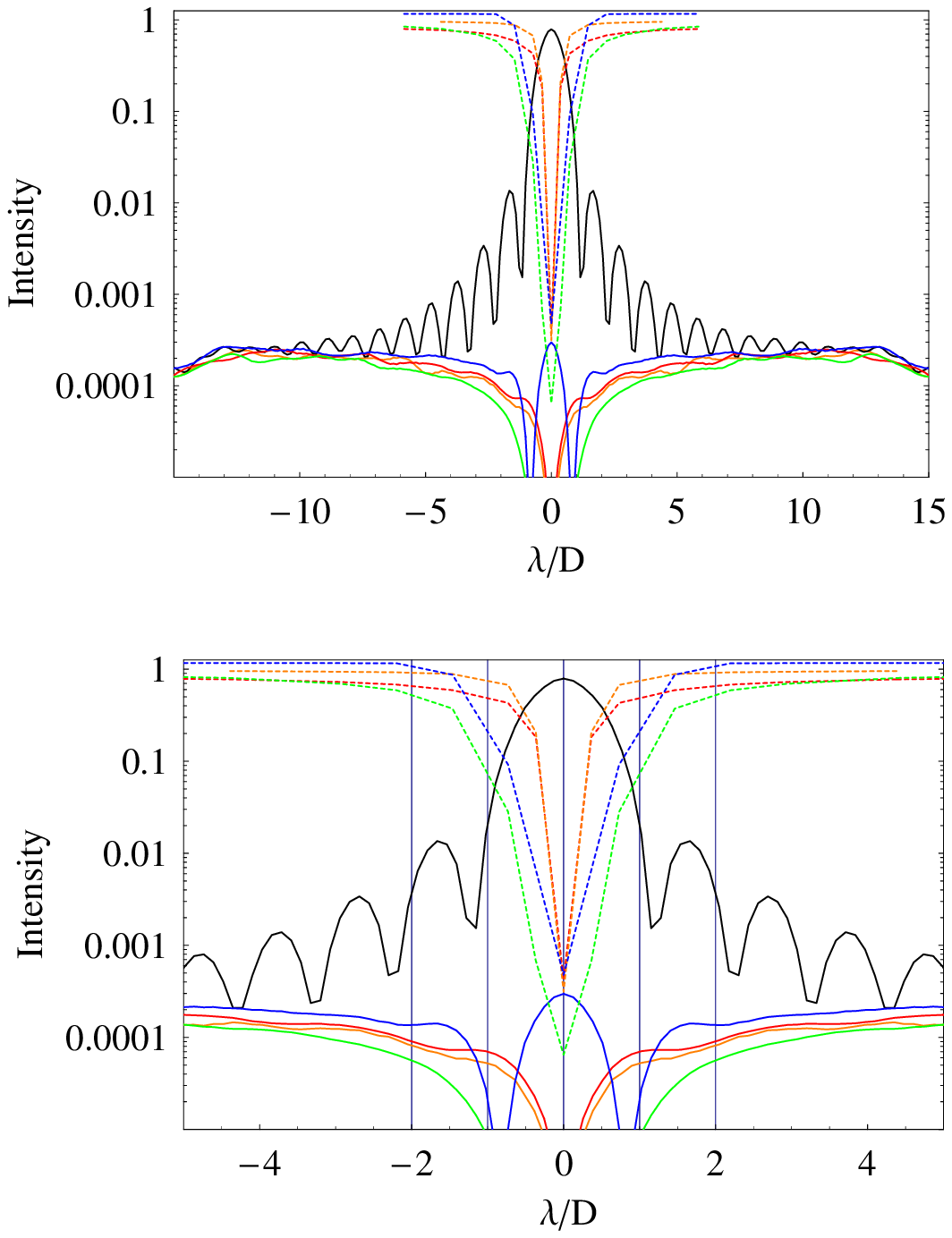}}
\caption{Simulation results for the long-exposure point spread
function of an on-axis source, for a well-corrected AO case (no
central obstruction, $2.5 d_{\mbox{eff}}=r_0$). The intensities are
normalized to the diffraction-limited case. Colours code the
no-vortex case $m=0$ (black), charges 2 (red) and 6 (green), as well
as the quadrant phase mask (orange) and prolate-apodized mask
(blue). Dashed lines show the relative throughput for an off-axis
source, as a function of off-axis distance.}\label{figure8}
\end{figure}
\clearpage

Each coronagraphic method works similarly outside a few $\lambda/D$.
As expected, the coherent core of the PSF is largely removed, and so
the attainable dynamic range is simply set by the strength of the
PSF halo at the shoulder. There are strong differences in
effectiveness within $\sim 2 \lambda/D$. These are ambitiously small
offsets to consider, since aberrations and scattering in the
telescope are extremely difficult to control at these levels.
Nonetheless, the ultimate purpose of coronagraphs is to be effective
at a few $\lambda/D$.

The question then arises of how well the various coronagraphs would
transmit an off-axis source.  What is the exclusion radius $r_c$
beyond which a source is unaffected by the coronagraph?

In the case of the OVC  $r_c$ can be estimated by requiring that the
phase introduced by the OVC (charge $m$), across the off-axis
source, should be less than about $\pi$.  Hence
\[ r_c = \frac{m}{\pi} \frac{\lambda}{D} \]
suggesting that high-charge OVCs will not be able to detect
companion sources as close as low-charge ones.  In fact, it is
possible to calculate analytically the response of an OVC to
off-axis sources in the diffraction limit (see Appendix 2) and this
shows that, close to on-axis, the response is a very strong function
of charge.  Larger charges attenuate very strongly close to the
axis.  Combining these two results suggests that overall higher
charges should have bigger exclusion radii and is a useful
analytical verification of the trends in the simulation results.

For the quadrant mask   a narrow radius of exclusion is expected,
roughly $r_c=(2/\pi)( \lambda/D)$ and rather like a charge 2 OVC.
The prolate case is too complicated to estimate because of
diffraction at a stop in the focal plane.

\begin{figure}[htb]
\center{\includegraphics[width=10cm]{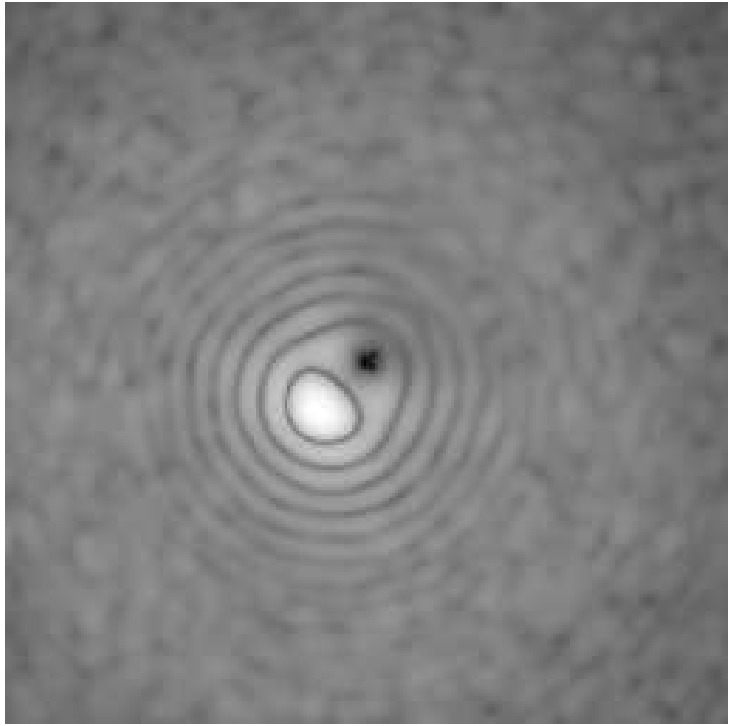}} \caption{A
gray-scale image illustrating a source emerging from the vortex in
the $m=6$ well=corrected case.}\label{figure9}
\end{figure}

As a source emerges ``from the vortex'' the PSF is quite distorted
(Figure \ref{figure9}).  To quantify the exclusion radius I used as
a metric the flux through a small circular aperture, extending out
to the first (diffraction limited) dark ring and centered at the
input position of the off-axis test source. This is normalized to
the flux through the same aperture for the partially-corrected PSF,
with no coronagraph.  The results for the four coronagraphs are
superimposed on Figures \ref{figure7} and \ref{figure8}.

\begin{figure}[htb]
\center{\includegraphics[width=.85\textwidth]{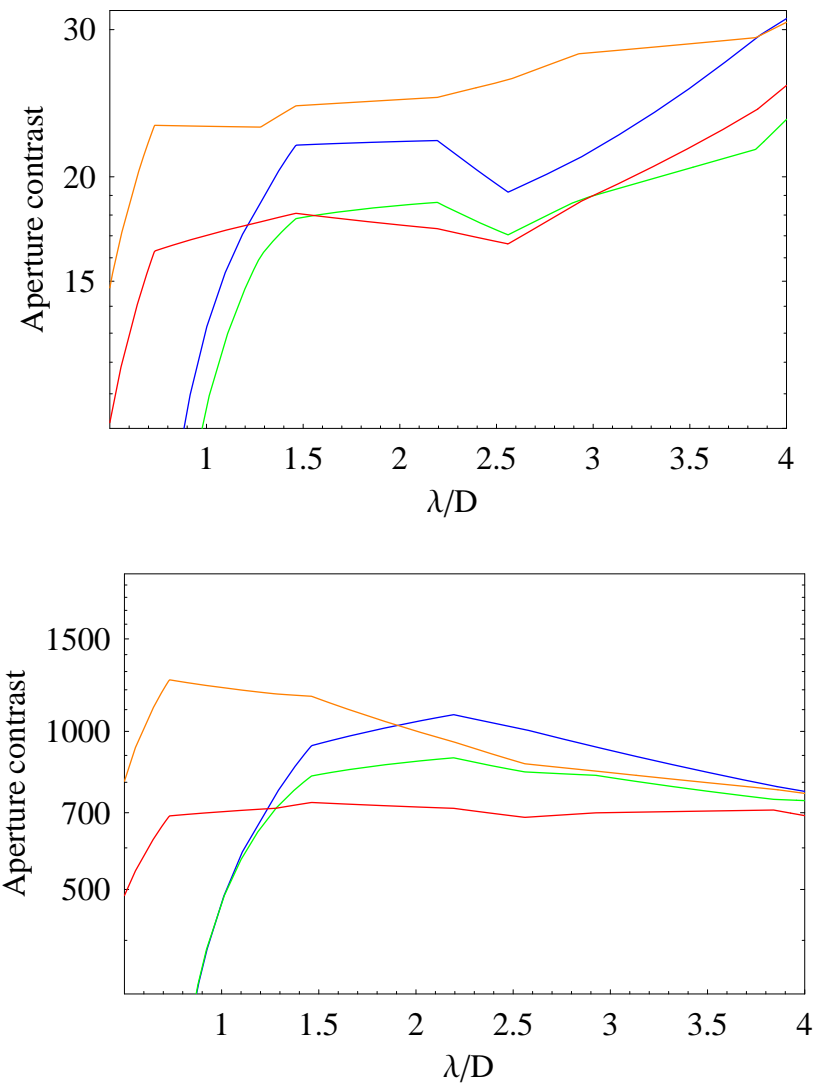}}
\caption{The attainable contrast through an aperture, as defined in
the text. The analytical results would suggest that performance
would converge (at least for the phase masks) at large radii and low
turbulence, as observed in these results.} \label{figure9a}
\end{figure}
\clearpage

Figure \ref{figure9a} plots the dynamic range, defined as the flux
through the aperture of an off-axis source, divided by the flux
though this aperture from the off-axis source, as attenuated by the
coronagraph.  The general trends are as expected; all designs
converge to a similar value at large radii, where performance is
fixed by AO parameters.  Inwards of this, as would be expected, the
charge 6 design does somewhat better than charge 2 until a point is
reached where the wider exclusion radius of the charge 6 mask
attenuates the off-axis source markedly.  These trends are rather
similar for both levels of AO correction, although of course the
attainable dynamic range is much better for a better-corrected
wavefront.  What is most apparent however is that the quadrant mask
does better than the two pure vortex masks, and seems to have a
well-balanced mixture of various charges to perform well at both
large and small radii. The performance however will depend on angle
relative to the phase discontinuities in the mask, which produce
``blind lines''.  The prolate-apodized mask also performs well if
the loss of light is not an issue.

A charge-2 OVC would be a good choice for an instrument which had no
angular dependence of its response in the focal plane, but is more
complex to manufacture than a quadrant phase mask.  Whether there is
an ``optimum'' mask of this type in some sense remains an open
question.

\section{Conclusions}

The phase vortex mask is an interesting basis for a coronagraph
because it offers the possibility of complete on-axis extinction of
a pure Airy diffraction pattern.  This result can be demonstrated
analytically for even topological charges. Moreover, masks of even
charges form the Fourier basis to build up many other varieties of
phase mask, for example the quadrant mask, all of which will share
the extinction property, and analytical tractability, of a
single-charge mask.  Odd charges always degrade performance so the
class of useful masks is restricted by this fact.

A ground-based telescope however will typically have two aspects
that degrade the performance of vortex phase masks, and phase masks
derived from them. The central obstruction of the telescope, and the
effects of atmospheric turbulence (even partially corrected) are
limitations.  ``Removing'' the central obstruction, by Guyon's
method for example, seems essential to attain anything more than
modest dynamic range, a conclusion which applies to either
diffraction-limited images or moderate levels of AO correction. This
of course is a common problem for coronagraphs.

The effect of this class of phase masks can be analyzed quite
generally when partially corrected turbulence is present, and the
conclusion is that they will tend to remove the coherent core of the
AO-corrected images and leave the incoherent power-law halo.  The
implied outer working radius is  the ``shoulder'' in a typical AO
image where the core meets the halo, largely independently of the
details of the mask but fixed by details of the AO system such as
the actuator density.

The inner working radius is however affected by the details of the
mask. It is determined by the extinction of a hypothetical source
adjacent to the on-axis one. Here the masks do differ, and it is
possible to show that high-charge masks have a wider zone of
extinction than low-charge ones. Taken together with the residual
light that is transmitted from the on-axis source, the detailed
simulations show that there is rather little to choose between the
various masks considered here. In fact the quadrant mask is as good
as any. High-charge masks seem to offer relatively small gains, for
the particular turbulence parameters that were examined. The
quadrant mask has the advantage of being relatively simple to
manufacture\cite{riaud}, by comparison with a vortex where a smooth
gradient in phase has to be achieved over the mask.  Overall however
the phase masks are not the limiting factor in dealing with AO
imagery and it seems likely that a suitable mask could be designed
for most applications by suitable tailoring of the basis set of
vortex masks.

\section*{Appendix 1: Effect of the vortex filter on a partially-corrected image}

In the Fourier optics approximation, as is well known, the
mathematical description of a coronagraph requires three Fourier
transforms.  The first operates in the exit pupil of the telescope
or adaptive optics system, and produces the amplitude distribution
in the focal plane. This amplitude is modified by a stop or phase
plate.  A further (inverse) transform gives the resultant amplitude
in a reimaged pupil plane, and here a Lyot or pupil stop is
typically used to eliminate unwanted diffraction.  A third transform
is used to obtain the  amplitude at a final focal plane.  I will
denote co-ordinates in the first pupil by $r$, the first focal plane
by $k$, in the second pupil plane by $r'$ and in the final focal
plane by $k'$.  A circular, unobstructed aperture is assumed.

For a well-corrected adaptive optics system, the amplitude in the
exit pupil is given by $\exp (\imath \phi(\vec{r})) \simeq 1+\imath
\phi(\vec{r})$.  If the number of actuators is large (the
mathematical requirement is that the scale of correlation of the
phase is much shorter than the pupil size) the a standard argument
gives the long-exposure point spread function (PSF).  This is the
diffraction-limited PSF of the pupil, scaled by the Strehl ratio,
plus a halo which is proportional to the power spectrum of the
phase, scaled by (1-Strehl).  (The appearance of a second-order
quantity, namely the power spectrum of the phase, indicates that for
consistency the phase exponential has to be expanded to second
order.  Carry these second-order terms through the calculation
gives the appropriate scalings of ``core'' and ``halo'' by phase
variance or Strehl ratio.)

Subject to the same assumptions, it is possible to find the PSF in
the final focal plane, after application of a vortex phase filter in
the first focal plane, and a pupil stop in the reimaged pupil.  For
an unobstructed diffraction-limited system, the final PSF is of
course zero; for partial correction, the PSF is just the phase power
spectrum. The vortex filter removes the coherent component, the Airy
disc, and only the incoherent part passes through.

I now give the details of this argument, which, although cumbersome,
involves only the repeated application of the physical
approximations already stated.  For simplicity, constants of
proportionality that are unnecessary to the main argument will be
omitted.

In the first focal plane, the  amplitude will be (proportional to is
understood from here on) -

\[{\cal A}(\vec{k}) = {\cal A}(\vec{k}) + \imath \int \vec{dr} \, \phi(\vec{r})
\exp (\imath \vec{k} \cdot \vec{r}) \Omega(\vec{r}) \]

in which $\vec{k}$ denotes a (suitably scaled) position in  the
focal plane, $A$ is the Airy amplitude pattern, and
$\Omega(\vec{r})$ is a function which is unity inside the pupil and
zero outside.

This amplitude is multiplied by a vortex filter function
$f(\vec{k})$ which is of the form $\exp(\imath m \theta)$, $m$ being
an even integer and $\theta$ being the azimuthal angle in the focal
plane.  $f f^* =1$ will be a useful property later in the argument.

Propagating to the reimaged focal plane, and applying the full-size
pupil stop there gives an amplitude leaving that pupil which is

\[
\Omega(\vec{r'}) \int \vec{dk} \, {\cal A}(\vec{k}) f(\vec{k})\exp
(-\imath \vec{k} \cdot \vec{r'}). \]

Here $\vec{r'}$ is the co-ordinate in the reimaged pupil.  The
extinction property of the even-charge vortex filter is that

\[\Omega(\vec{r'}) \int \vec{dk} \,A(\vec{k}) f(\vec{k})\exp (-\imath
\vec{k} \cdot \vec{r'})=0, \]

so the amplitude leaving the reimaged pupil is

\[
\imath \Omega(\vec{r'}) \int \vec{dk} \, \left( \int \vec{dr} \,
\phi(\vec{r}) \exp (\imath \vec{k} \cdot \vec{r}) \Omega(\vec{r})
\right) f(\vec{k})\exp (-\imath \vec{k} \cdot \vec{r'}). \]

Finally, therefore, the amplitude in the second focal plane will be

\[
{\cal A}'(\vec{k'})=\int \vec{dr'} \,  \imath \Omega(\vec{r'})
\int \vec{dk} \,  \int \vec{dr} \, \phi(\vec{r}) \exp (\imath
\vec{k} \cdot \vec{r}) \Omega(\vec{r})  f(\vec{k})\exp
(-\imath \vec{k} \cdot \vec{r'})  \exp (\imath \vec{k'} \cdot
\vec{r'}).
\]
The time- or ensemble-averaged intensity is

\[ I(\vec{k'})=\left< {\cal A}'(\vec{k'}){\cal A}'^*(\vec{k'}) \right>. \]

To carry out the sixfold integration and averaging,  write

\begin{eqnarray}
{\cal A}'^*(\vec{k'}) & = & \int \vec{dr_1'} \,  (-\imath)
\Omega(\vec{r_1'}) \int \vec{dk_1} \,  \int \vec{dr_1} \,
\phi(\vec{r_1}) \exp (-\imath \vec{k_1} \cdot \vec{r_1})
\Omega(\vec{r_1})  f(\vec{k_1})^*  \times \nonumber \\
& \times & \exp (\imath \vec{k_1} \cdot \vec{r_1'})  \exp (-\imath \vec{k_1'}
\cdot \vec{r_1'}). \nonumber
\end{eqnarray}

in which the subscript $1$ is used to indicate corresponding dummy
variables of integration.

Re-arranging the order of integration, the first part of the sixfold
integration involves $\vec{r}$ and $\vec{r_1}$:

\[ \int \int \vec{d r}\, \vec{dr_1} \, \left< \phi(\vec{r})\phi(\vec{r_1}) \right>
\exp\left(\imath \vec{k}\cdot \vec{r} - \imath \vec{k_1}\cdot
\vec{r_1}\right) \Omega(\vec{r}) \Omega(\vec{r_1}).
\]

Writing $\vec{r_1}=\vec{r}+\vec{u}$, the integral contains the term

\[ \left< \phi(\vec{r})\phi(\vec{r}+\vec{u}) \right> \simeq {\cal C}(\vec{u})\]

which is approximately equal to the spatially-invariant phase
autocorrelation function ${\cal C}$ (the equality cannot be exact
because of the edges of the pupil).  If ${\cal C}$ declines rapidly
with $\vec{u}$ (many actuators), then the $\vec{u}$ and $\vec{r}$
integrals separate to give the approximate result for this first
pair of integrals as

\[ {\cal P}(\vec{k_1}) A(\vec{k}-\vec{k_1})\]

in which $P$ is the phase power spectrum.  This  will turn out to be all that
passes through the vortex coronagraph.

The next pair of integrals is

\[ \int \int \vec{dk} \vec{dk_1} f(\vec{k})  f(\vec{k_1})^*
\exp\left(-\imath \vec{k}\cdot \vec{r'} + \imath \vec{k_1}\cdot
\vec{r_1'}\right) {\cal P}(\vec{k_1}) A(\vec{k}-\vec{k_1}).
\]
The Airy amplitude $A$ is strongly peaked compared to other terms,
and picks out $\vec{k}=\vec{k_1}$.  The phase function $f$ has unit
modulus and so disappears, leaving

\[ \int \vec{dk} \exp\left(-\imath \vec{k}\cdot \vec{r'} + \imath \vec{k}\cdot \vec{r_1'}\right){\cal P}(\vec{k})
= {\cal C}(\vec{r_1'}-\vec{r_1}).\]

The final pair of integrals is

\[\int \int \vec{dr} \vec{dr_1}
\exp\left(-\imath \vec{k'}\cdot \vec{r'} - \imath \vec{k'}\cdot
\vec{r_1'}\right) {\cal C}(\vec{r_1'}-\vec{r_1})
\Omega(\vec{r'})\Omega(\vec{r_1'}).
\]
With the same change of variable, and the same argument about the
scale of ${\cal C}$ compared to the pupil size, it follows that

\[I(\vec{k'})\simeq P(\vec{k'})
\]
so that the effect of the vortex filter is to remove the coherent
Airy core of the PSF and leave only the incoherent halo.

\section*{Appendix 2: Analytical estimate of off-axis behaviour of a vortex}
In the high-Strehl approximation the optical vortex filter acts
mainly on the diffraction-limited core of the AO-corrected PSF. It
is important to know what happens when this core is not centered on
the vortex filter, as this will be the case for point sources of
interest near to a bright star.  In the high-Strehl limit, where the
amplitude is linear in the residual phase,  the filter will act on
the off-centered Airy disc but  pass the uncorrelated (halo) light,
as is the case for an on-axis source.  In this section I show that
the transmitted energy is a simple power-law function of the offset
distance in the focal plane.

Let us therefore consider an Airy pattern, off-axis by an amount $s$
at some angle $\alpha$ (with respect to the origin of the focal
plane co-ordinate system $(k,\theta)$. To simplify the discussion, I
work in scaled units of length in both the pupil plane and focal
plane. In the pupil plane,  scale lengths by the pupil radius, and
in the focal plane,  scale by the factor $F \lambda/\pi$, $F$ being
the focal ratio of the system and $\lambda$ the operating
wavelength. This means that the Fourier transforms and PSFs can be
written compactly.

The functional form of the off-axis Airy pattern is the standard one
in a variable $\kappa$, with

\[ \kappa=\left( k^2+s^2- 2 k s \cos (\alpha -\theta)  \right)^{1/2} \]

$k$ being the scaled radial co-ordinate in the focal plane system
centered on the vortex filter.  The Airy pattern is therefore given
(to within  constants irrelevant to this discussion) by
$J_1(\kappa)/\kappa$.

To write this in the needed variables, namely $k$ and $s$, I apply
the Gegenbauer identity\cite{watsonb}. Applying this to the present
case,

\begin{equation}
  \frac{J_1(\kappa)}{\kappa} = 2
\sum_{n=0}^{\infty}(1+n)\frac{J_{n+1}(s)}{s} \frac{J_{n+1}(k)}{k}
C^{1}_n(\cos (\alpha-\theta)) \label{gegenbauer}
\end{equation}

in which $C^{1}_k$ is a Gegenbauer polynomial\cite{Abramowitz}.

To evaluate the transmission of an off-axis source through the
centered vortex filter, we can perform a Fourier transform, term by
term, of the amplitude after the vortex filter

\[ \exp  (\imath m \theta) \, \frac{J_1(\kappa)}{\kappa} \]

in which the transform is with respect to the variables $k$ and
$\theta$, and carries the intensity into a re-imaged focal plane in
which the co-ordinates are $\vec{r}$.

If the source is on axis, there is no amplitude within the re-imaged
pupil $\Omega(\vec{r})$ - this is the coronagraphic behaviour of the
vortex filter. Off-axis, we can find the leading order term in $s$
for which there is a finite amplitude within $\Omega$.  This gives
an estimate of the near-to-axis transmission behaviour of the vortex
filter.

Off axis (non-zero $s$) is more complicated mathematically but
involves no new physics.  The steps are as follows.

(1) In the Fourier transform into the re-imaged pupil

\[ \int d\vec{k} \, \exp  \imath m \theta \, \frac{J_1(\kappa)}{\kappa} \,
\exp\left(\imath \vec{k}\cdot\vec{r} \right)\]

we note that

\[ \vec{k}\cdot\vec{r} = k r \cos(\beta-\theta) \]

in which the polar co-ordinates in the re-imaged pupil are $r$ and
$\beta$.  The transform is therefore

\[ \int k \, dk \int d \theta \, \exp  (\imath m \theta ) \,
\frac{J_1(\kappa)}{\kappa} \, \exp\left( \imath k r
\cos(\beta-\theta)\right)
\]

(2) We then insert the Gegenbauer expansion for
$J_1(\kappa)/\kappa$.  Expanding the Gegenbauer polynomials, and
using multiple-angle formulas, rather than powers, we find that the
angular integral in the transform involves terms which are of the
 form

\[\int d\theta \, \exp \left( \imath m \theta\right) \exp\left( \imath k r
\cos(\beta-\theta)\right) \cos(p(\alpha-\theta)) \]

with $m$ and $p$ both integers.  These integrals can be
systematically evaluated by using the identity\cite{watsonc}

\begin{equation}
 \exp\left( \frac{z}{2}\,(t-\frac{1}{t})
\right)=\sum_{n=-\infty}^\infty t^n J_n(z), \label{generating}
\end{equation}

with the substitution $t=\exp(\imath(\beta-\theta))$, followed by
multiplication by the appropriate cosine factor and then
term-by-term integration.  The result involves terms in $J_{p-m}$
and $J_{p+m}$, as well as trigonometrical factors in $\beta$.

(3) The purpose of this calculation is to evaluate the transmission
of the coronagraph at small $s$, so at this stage we limit the
Gegenbauer series by expanding the $J_{n+1}(s)/s$ term to third
order in $s$.  This limits the expansion to fourth order Bessel
functions and a third order Gegenbauer polynomial.

(4)  We now have the $k$-integral to do.  This resolves into a
number of integrals of the Weber-Schafheitlin type (\cite{watsona},
namely

\[ \int_0^\infty \frac{J_\mu(a t)J_\nu(bt)}{t^\lambda} \, dt \]
which have a discontinuity at $a=b$ and different behaviour at $a>b$
and $a<b$.  In our case, this behaviour corresponds to whether any
light is diffracted beyond the vortex filter into the re-imaged
pupil.

(5)  We carried out this lengthy series of calculations in the
computer algebra package {\em MATHEMATICA}, and investigated
topological charges on the vortex masks of $m=2,3,4,5,6$.  What is
of interest is to note the first power of $s$, the focal-plane
offset, that gives any diffracted light {\em within} the re-imaged
pupil. If this power is $\gamma$ then this means that the
transmission energy of an off-centered image is proportional to
$(s/s_0)^{2 \gamma}$.  For odd topological charges, light is
diffracted into the pupil regardless of offset; this corresponds to
the known result that these charges are ineffective in a
coronagraph.  For even charges, we find the interesting result that
$2\gamma=m$, so that the exponent equals the charge. This means that
high-charge vortex filters will have a much steeper response to
off-axis objects.

(6) Finally, we evaluate $s_0$ by integrating the leading-order term
in $s$ over the re-imaged pupil.  What is of interest here is the
relative behaviour, so we evaluate with respect to the energy passed
by a $m=0$ filter. This takes care of the various multiplicative
constants that have been ignored so far.  In (focal plane) scaled
variables, the transmission of energy $I$  for charges 2, 4 and 6 is
given by
\begin{eqnarray}
I_2 & = & \frac{s^2}{6}   \nonumber\\
I_4 & = & \frac{s^4}{32}  \nonumber \\
I_6 & = & \frac{s^6}{240} \nonumber
\end{eqnarray}

This result is illustrated in Figure \ref{figure10}

\begin{figure}[htb]
\center{\includegraphics[width=10cm]{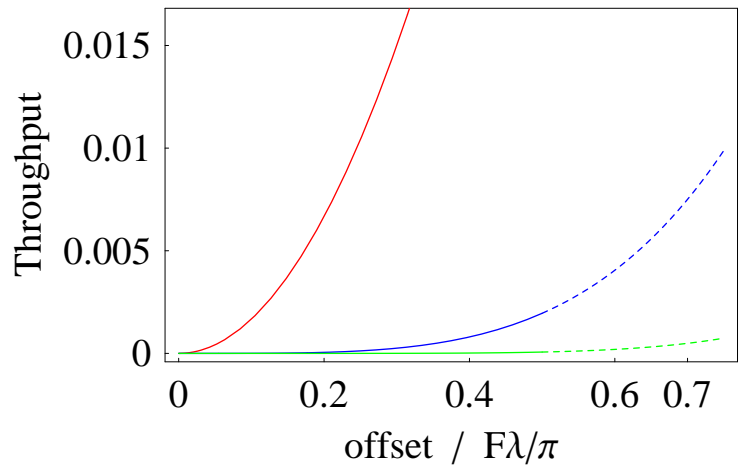}} \caption{The
transmitted intensity through a vortex, as a function of off-axis
angle, for charges 2 (red), 4 (blue) and 6 (green).  The analytical
approximation only applies at small $s$ and to emphasize this, lines
are shown dashed beyond $s=0.5$} \label{figure10}
\end{figure}
\clearpage

Here is a specific example of the calculation for a charge 2 vortex.
Expanding the Gegenbauer series (Equation \ref{gegenbauer}) to third
order in $s$ gives

\begin{eqnarray}
\frac{J_1(\kappa)}{\kappa} & = &  \frac{J_1(k)}{k}+\nonumber\\
& + & s \left(\frac{ J_2(k) \cos (\theta-\alpha )}{r}\right)+  \nonumber\\
 &+& {s^2}\left(\frac{ -3 J_1(k)+3
J_3(k)+6 J_3(k) \cos
(2 \theta-2 \alpha )}{24 k}\right)+  \nonumber\\
 &+& {s^3}\left(\frac{ -2 J_2(k)
\cos (\theta-\alpha)+J_4(k) \cos (\theta-\alpha )+J_4(k) \cos (3
\theta -3 \alpha)}{24 k}\right) \nonumber
\end{eqnarray}

The Fourier transform back into the pupil involves  the term-by-term
integration

\[ \int k \, dk \int d \theta \, \exp  (\imath 2 \theta ) \,
\frac{J_1(\kappa)}{\kappa} \, \exp\left( \imath k r
\cos(\beta-\theta)\right)
\]

in which we specialize to the charge 2 case as an illustration.  The
variable $r$ is the scaled co-ordinate in the re-imaged pupil plane.

The first term (no dependence on offset $s$) for the pupil-plane
amplitude $A$ is

\[
\begin{array}{lccl}
 A(r) & = & 0 &    0 \leq r < 1 \\ \\
& = & \displaystyle\frac{-2\pi \exp(2 \imath \beta)}{r^2}  & r \geq 1.\\
\end{array}
\]

The next term (linear in $s$) is
\[
\begin{array}{lccl}
 A(r) & = & \imath \pi s \exp(\imath(\alpha+\beta))\, r  &    0 \leq r < 1
 \\ \\
& = & \displaystyle \frac{-\imath \pi s \exp( -\imath (\alpha- 3\beta))}{r^3}  & r \geq 1.\\
\end{array}
\]

Here we see the first appearance of  finite amplitude {\em within}
the pupil.   For larger charges the calculations have to be carried
to higher in $s$ before this happens.  Figure \ref{figure11}
illustrates the appearance of the reimaged pupil, calculated in this
analytical fashion, for a small offset of a source in the case of a
charge 6 vortex.

\begin{figure}[htb]
\center{\includegraphics[width=10cm]{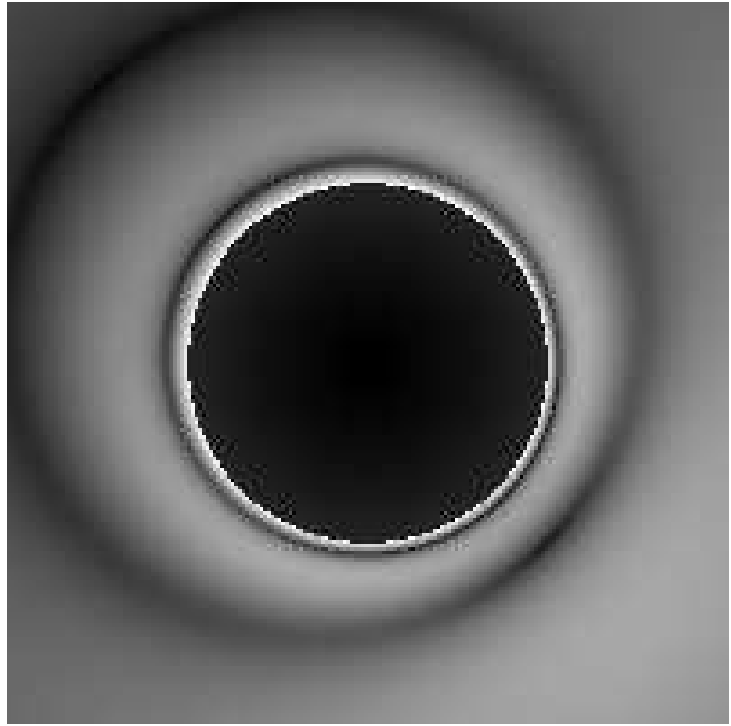}} \caption{The
transmitted intensity (stretched scale) in the pupil through a
charge 6 vortex, for a small offset of the source in the focal
plane.  This was calculated analytically by the method in the text.}
\label{figure11}
\end{figure}

A useful check is to repeat this calculation for no vortex, namely a
charge of zero.  Although complicated expressions result, they
simplify as expected; the intensity is uniform over the pupil, with
a constant phase gradient.

\section*{Appendix 3.  The quadrant mask as a superposition of vortex
masks}

The quadrant mask amplitude is defined as a function of azimuthal
angle $\theta$ in the focal plane as

\begin{displaymath}
\begin{array}{lcll}
{\cal A}(\theta) & = & 1 &   0 \leq \theta < \pi/2 \\
& & &  \\
{\cal A}(\theta) & = & -1 &   \pi/2 \leq \theta < \pi \\
& & &  \\
{\cal A}(\theta) & = & 1 &   \pi \leq \theta < 3\pi/2 \\
& & &  \\
{\cal A}(\theta) & = & -1 &   3\pi/2 \leq \theta < 2\pi \\
& & &  \\
\end{array}
\end{displaymath}
with no radial dependence.  This amplitude can be expressed as a
Fourier series:

\begin{equation}
{\cal A}(\theta) = \frac{\pi}{2 \imath}(\ldots
+\frac{1}{5}e^{-10\imath \theta}+\frac{1}{3}e^{-6\imath \theta}
+e^{-2 \imath \theta}+ e^{2 \imath \theta}+ \frac{1}{3}e^{6\imath
\theta} +\frac{1}{5}e^{10\imath \theta}+\ldots) \label{fourier}
\end{equation}

so, a sum of various evenly charged vortex phase masks.

The effect of this (or other phase masks represented as Fourier
series) can be analyzed analytically for an Airy diffraction
pattern. Diffraction from the focal plane will involve integrals of
the form

\[
\int_0^\infty \, dr \, r \frac{J_1(k)}{k} \int_0^{2\pi}\, d\theta \,
{\cal A}(\theta) \exp(\imath k r \cos(\theta-\beta))
\]

which requires the result

\[ \int_0^{2\pi}\, d\theta \,
\exp (\imath n \theta) \exp(\imath z \cos(\theta-\beta))=2\pi \exp
(\imath n \beta) J_n(z)
\]

for $z$ real and $n$ an integer. (This can be deduced from Equation
\ref{generating} by the method described there.)

The radial integrals are then of the Weber-Schafheitlin form, so the
amplitude in the reimaged pupil is a weighted sum of the amplitudes
for the various vortex masks that appear in Equation \ref{fourier}.
Each of these is zero within the pupil as long as only evenly
charged vortices appear in Equation \ref{fourier}.

As an example, Figure \ref{figure12} shows a numerical and an
analytical calculation of the reimaged pupil plane after application
of a quadrant phase mask.  This technique allows analytical
calculations to be performed for any useful phase mask which does
not have a radial dependence of the phase.

\begin{figure}[htb]
\center{\includegraphics[width=10cm]{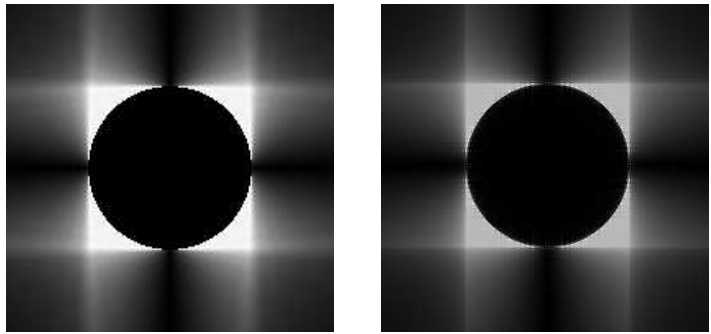}} \caption{The
transmitted intensity (stretched scale) in the pupil through a
quadrant phase mask, calculated numerically (right) and analytically
with ten non-zero terms in the Fourier expansion.} \label{figure12}
\end{figure}

\end{document}